# Non-exponential time-correlation function for random physical processes

T.R.S. Prasanna

Department of Metallurgical Engineering and Materials Science

Indian Institute of Technology, Bombay

Mumbai – 400076 India

The exponential correlation function is theoretically incorrect in the entire frequency range of interest for processes described in terms of linear response theory. The Lorentzian lineshape results from an inconsistent assumption of exponential correlation at timescales smaller than the relaxation time. A new correlation function is proposed that avoids the deficiencies of the exponential function. Comparison on dielectric relaxation in gases shows that the new correlation function can be used to fit data satisfactorily instead of the exponential function. The new correlation function is theoretically consistent for all processes described in terms of linear response theory. Its additional mathematical superiority implies that it can be used instead of the exponential function for all such processes.

The exponential time-correlation function is used widely to describe relaxation in many random physical processes in science and engineering. It appears to have been introduced by Debye [1] to represent dielectric relaxation in liquids and gases and has been used extensively ever since. Dielectric relaxation in gases belongs to a class of problems that are described in the framework of linear response theory [2] where the use of the exponential correlation function is widespread. However, this function is non-analytic at t=0 and results in several difficulties, including divergence of physical quantities. To overcome this problem, it is approximated at infinitesimal times by a Gaussian function that is continuous at t = 0. Hence a "short time approximation" is made near t=0 where the correlation function is a Gaussian and a "long term approximation" is made where it behaves as an exponential function [2,3]. However, in practice, these approximations are frequently ignored and the exponential function is used at all timescales. The exponential function is of theoretical importance as well, because according to Doob's theorem [4, 5], any stationary, Gaussian, Markovian process can only have an exponential time correlation function. Conversely, any stationary, Gaussian process with exponential correlation is also Markovian [4,5] for which a large body of knowledge exists [5-7].

In this paper, it is shown that the exponential correlation function is theoretically inconsistent in the entire frequency range of interest for processes described in the framework of linear response theory. This is independent of the microscopic mechanisms of relaxation that may be different for different phenomena. A new one-parameter function that avoids the theoretical and mathematical difficulties posed by the exponential function is proposed to describe relaxation phenomenologically. The results obtained from the new correlation function are compared with those from the exponential function using the example of dielectric relaxation in gases [2,11,12].

In linear response theory, the generalized displacement, B(t), is given by the convolution, $B(t) = \int_0^\infty \chi(\tau) F(t-\tau) d\tau$, of the response function, χ(τ), with the generalized force, F(t), that also incorporates the lag of the displacement and is non-Markovian [2]. Only when the force varies slowly is it given by $B(t) = F(t)\left(\int_0^\infty \chi(\tau) d\tau\right)$ that is Markovian [2]. The complex susceptibility [2] is given by $\chi(\omega) = \chi'(\omega) + i\chi''(\omega) = \int_0^\infty \chi(\tau) e^{i\omega\tau} d\tau$. The process becomes non-Markovian at frequencies where the lag δ, given by tan δ = χ''(ω) / χ'(ω), acquires a finite value. This occurs when χ''(ω) starts rising from its zero value at low frequencies (see also fig.3). Below this frequency, the static susceptibility, χ'(0), is sufficient to characterize the response completely. Therefore, whenever χ''(ω) is of interest, the random process is non-Markovian but using an exponential function implies that the process is Markovian [4,5]. Hence, the exponential correlation function is *theoretically incorrect in the entire frequency range of interest* even though it may provide a satisfactory fit to experimental data. The exponential correlation function, $C_1(\tau) = e^{-|\tau|/\tau_0}$, is obtained when linear relaxation, $\frac{d}{dt}\langle X(t)\rangle = -\frac{1}{\tau_0}\langle X(t)\rangle$ is assumed [2]. The above discussion shows that this expression is flawed in the entire frequency range (or timescale) of interest for many random processes.

This is consistent with the fact that most random processes are non-Markovian and can be approximated as Markovian only after a "coarse graining" so that the timescales of interest are much larger than the relaxation time. When the timescale of interest is of the same order as the relaxation time, the process cannot be approximated as Markovian [2].

A more general argument (valid beyond linear response theory) is that an exponential correlation function is inconsistent whenever the derivatives of random processes have physical meaning (velocity fluctuations in fluids, current fluctuations in inductors etc.) as it implies that the process is not mean-square differentiable [3].

From the above discussion, it is clear that a theoretically consistent, i.e. non-exponential, time-correlation function needs to be developed. One approach is to represent the microscopic physics correctly in which case the correlation function will be specific to individual phenomena. Another is to adopt a phenomenological approach that will provide a first approximation to time-correlation for many random processes. Because the exponential function represents the second approach, the new correlation function is developed in the same framework.

The properties a correlation function for any stationary random process, $X(t)$, must possess [3,13] are a) $C(-\tau) = C(\tau)$ and b) $|C(\tau)| \leq C(0)$. Also, if the derivatives have physical meaning, additional properties it must possess [3,13] are c) $C(\tau)$ must be continuous at $\tau = 0$ for the process to be mean-square continuous, d) $C(\tau)$ must be twice differentiable for the derivative of a random process, $X'(t)$, to exist and $<[X'(t)]^2> = -$

C″(0) and e) the random process and its derivative are uncorrelated at the same instant, C′(0) = <X(t) X′(t)> = 0 [3,13]. The power spectrum, S(ω), is the Fourier Transform of the time-correlation function by the Wiener-Khinchin theorem [11,13]. Expanding the time-correlation function, C(τ), about τ = 0 in a Taylor's series and expressing the derivatives in terms of integrals over S(ω) results in an expression

$$C(\tau) = \frac{1}{2\pi} \sum_{n=0}^{\infty} \frac{(i\tau)^n}{n!} \int_{-\infty}^{\infty} \omega^n S(\omega) d\omega \qquad (1)$$

that describes the time-correlation function in terms of moments of the power spectrum. This expansion does not exist for the exponential correlation function as it is non-analytic at τ = 0. If the dynamical variables of the system determine the correlation function, it must be analytic at t = 0 [2] which implies that its power spectrum must have finite moments. Eqn.1 suggests one criterion for selecting new correlation functions i.e. all moments of the power spectrum must be finite. S(ω) is an even function as classical time-correlation functions are real and only even powers contribute to the summation in eqn.1. One of the simplest functions for which all moments are finite is the exponential function, $S(\omega) = e^{-\alpha|\omega|}$. In this case, the various moment integrals become gamma functions and are given by $\frac{2}{\alpha^{n+1}} \Gamma(n+1)$. This corresponds to a time-correlation function of the form $a/a^2+\tau^2$ and when normalized to give C(0) = 1 can be written as

$$C_2(\tau) = \frac{a^2}{a^2 + \tau^2} \qquad (2)$$

Fig.1 plots this function (a = 1.4 T) along with the exponential function, $C_1(\tau) = e^{-|\tau|/\tau_0}$ with $\tau_0$ = 2 T. The new function requires only one parameter (a) to be specified like the exponential function ($\tau_0$) and does not differ substantially numerically from it. Another feature is that it possesses an inflection point at finite time $\tau = a$, which corresponds to the mean collision time when relaxation occurs by molecular collisions. This is because the rate of change of correlation must be maximum at the mean collision time leading to an inflection point, $d^2C/d\tau^2 = 0$, at this time. The mean collision ($\tau_0 = 2$ T, a = 1.4 T) and relaxation ($\tau_0 = 2$ T, $\tau_2 = 2.2$ T) times are comparable. The exponential time-correlation function does not satisfy conditions d) and e). The new function is analytic at $\tau = 0$ and satisfies all conditions, a) – e), required of a correlation function [3,13] and hence is mathematically superior to the exponential function. It will also satisfy various sum rules [2] as all moments of its power spectrum are finite.

The results from the two correlation functions for the problem of dielectric relaxation [2,11,12] in gases are compared. The Debye model assumes an exponential correlation function. It does not account for molecular rotations and does not possess an inflection point that must exist at the mean collision time if relaxation is by collisions of molecules. Accounting for molecular rotations by a $\cos\omega_0\tau$ factor results in a correlation function

$$C_3(\tau) = \mu_{z0}^2 \, e^{-|\tau|/\tau_0} \cos\omega_0\tau \tag{3}$$

proposed by Van Vleck-Weisskopf [8] and Frohlich [9] that has the same theoretical deficiencies as the exponential function. A new correlation function that takes into account molecular rotation and is theoretically consistent can be constructed as

$$C_4(\tau) = \mu_{z0}^2 \, \frac{a^2}{a^2 + \tau^2} \cos\omega_0\tau \tag{4}$$

Their Fourier transforms are the lineshape functions for absorption of radiation. They are the Lorentzian lineshape [11] (for $\omega > 0$)

$$I_3(\omega) = \frac{1}{2\pi}\int_{-\infty}^{\infty} C_3(\tau)e^{-i\omega\tau}d\tau \approx \frac{\mu_{z0}^2}{2\pi}\left[\frac{1/\tau_0}{\frac{1}{\tau_0^2}+(\omega-\omega_0)^2}\right] \quad (5)$$

and a new lineshape function obtained by referring to tables [14]

$$I_4(\omega) = \frac{1}{2\pi}\int_{-\infty}^{\infty} C_4(\tau)e^{-i\omega\tau}d\tau = \frac{\mu_{z0}^2 a}{2}\left[\begin{array}{ll} e^{-\omega_0 a}\cosh a\omega & (\omega \leq \omega_0) \\ e^{-\omega a}\cosh a\omega_0 & (\omega \geq \omega_0) \end{array}\right] \quad (6)$$

Fig.2 plots the lineshape functions and shows that they are practically indistinguishable. In eqn. 5, the frequencies ω range about $\omega_0$ that is greater than the "relaxation frequency" $\omega_c = 1/\tau_0$. This suggests that an exponential correlation has been assumed at timescales smaller than the relaxation time, $\tau_0$. Therefore, the Lorentzian lineshape results from an incorrect theoretical basis. Additionally, it has divergent moments [11]. The new lineshape is theoretically consistent and has finite moments to all orders, which suggests that it can be used instead of the Lorentzian lineshape.

The dielectric constants [2,11] for the correlation functions $C_1(\tau)$ to $C_4(\tau)$ obtained by referring to tables [14-16] are given in Table 1. The constant k accounts for the slightly different expressions also found in literature [11]. The real parts of the new dielectric constants, $\varepsilon_2'(\omega)$ and $\varepsilon_4'(\omega)$, must be evaluated numerically.

Figs.3 and 4 plot the imaginary parts of the dielectric constants $\varepsilon_1''(\omega)$, $\varepsilon_2''(\omega)$ and $\varepsilon_3''(\omega)$, $\varepsilon_4''(\omega)$ respectively. Above $\log(\omega) \approx -2.0$ T$^{-1}$, $\varepsilon_1''(\omega)$ and $\varepsilon_2''(\omega)$ start rising from a zero value (Fig.3) implying that above this frequency the polarization begins to lag behind the field. Hence, beyond this frequency the process is non-Markovian and $\varepsilon_1''(\omega)$, from the Debye model, is *theoretically incorrect* (though it may provide a satisfactory fit to the data) while $\varepsilon_2''(\omega)$ is not. This differs significantly from the current view that the Debye model is adequate except at high frequencies where new mechanisms become operative [2,11,12]. In Fig4, $\varepsilon_3''(\omega)$, from the Van Vleck-Weisskopf model, is theoretically incorrect while $\varepsilon_4''(\omega)$ is not. The total power loss given by $\int_0^\infty \omega \varepsilon''(\omega) d\omega$ diverges in the Debye and Van Vleck-Weisskopf models [2] but is finite for the new functions $\varepsilon_2''(\omega)$ and $\varepsilon_4''(\omega)$. Hence, from figs.1-4, it can be concluded that the new correlation function provides a satisfactory numerical alternative to the exponential function. In addition, it is theoretically consistent and mathematically superior to the exponential function.

In conclusion, the exponential correlation function is a) theoretically incorrect in the entire frequency range of interest for random processes described in terms of linear response theory and b) is, more generally, incompatible for all processes where the derivatives have physical meaning. Its use results in inconsistencies such as a) a Lorentzian lineshape that implies an exponential correlation at timescales smaller than the relaxation time and b) infinite power loss. A new one-parameter correlation function is proposed as a mathematically superior alternative to fit experimental data satisfactorily. It is theoretically consistent for the above-mentioned categories of random processes and can be used instead of the exponential function for all such processes.

1. P. Debye, *Polar Molecules,* (Chemical Catalog Co., New York, 1929)
2. R. Kubo, M. Toda and N. Hashitsume, *Statistical Physics II Nonequilibrium Statistical Mechanics,* (Springer, Berlin 1991)

3. P. Beckmann, *Probability in Communication Engineering,* (Harcourt, New York 1967)

4. J. L. Doob, Annals Math. **43** 351 (1942)

5. N. G. Van Kampen, *Stochastic Processes in Physics and Chemistry,* (North-Holland, Amsterdam 1981)

6. A. T. Bharucha-Reid, *Elements of the Theory of Markov Processes and their Applications,* (McGraw-Hill, New York 1960)

7. M. Sharpe, *General Theory of Markov Processes,* (Academic, Boston 1988)

8. J. H. Van Vleck and V. H. Weisskopf, Rev. Mod. Phys. **17**, 227 (1945)

9. H. Frohlich, *Theory of Dielectric Constant and Dielectric Loss*, (Oxford, Oxford 1949)

10. E. P. Gross, Phys. Rev. **97** 395 (1955)

11. D. A. McQuarrie, *Statistical Mechanics,* (Harper, New York 1976)

12. M.W. Evans, G. J. Evans, W. T. Coffey and P. Grigoline, *Molecular Dynamics and Theory of Broad Band Spectroscopy,* (Wiley-Interscience, New York 1982)

13. A. M. Yaglom, *Correlation Theory of Stationary and Related Random Functions I: Basic Results,* (Springer-Verlag, New York 1987)

14. F. Oberhettinger, *Tables of Fourier Transforms and Fourier Transforms of Distributions,* (Springer, Berlin 1990)

15. A. Jeffrey, *Handbook of Mathematical Formulas and Integrals,* (Academic, San Diego 2000)

16. I. S. Gradshteyn and I. M. Ryzhik, *Tables of Integrals, Series and Products,* (Academic, Boston 1990)

Table I   Real (ε') and imaginary (ε") parts of the dielectric constant for various correlation functions, $C_1(\tau)$ to $C_4(\tau)$. $\varepsilon_1$, $\varepsilon_3$ are from the Debye, Van Vleck-Weisskopf models and $\varepsilon_2$, $\varepsilon_4$ are their respective equivalents developed in this paper.

| $\varepsilon'(\omega) - \varepsilon_\infty = k\int_0^\infty -(dC(\tau)/d\tau)\cos\omega\tau\, d\tau$ | $\varepsilon''(\omega) = k\int_0^\infty -(dC(\tau)/d\tau)\sin\omega\tau\, d\tau$ |
|---|---|
| $\varepsilon_1'(\omega) - \varepsilon_\infty = k\,(1/1+\omega^2\tau_0^2)$ | $\varepsilon_1''(\omega) = k\,(\omega\tau_0/1+\omega^2\tau_0^2)$ |
| $\varepsilon_2'(\omega) - \varepsilon_\infty =$ $k + k\dfrac{\omega a}{2}\left(e^{-a\omega}\overline{E}i(a\omega) - e^{a\omega}Ei(-a\omega)\right)$ | $\varepsilon_2''(\omega) = k(\pi a/2)\omega\, e^{-a\omega}$ |
| $\varepsilon_3'(\omega) - \varepsilon_\infty =$ $\dfrac{k}{2}\left(\dfrac{1+\omega_0(\omega+\omega_0)\tau_0^2}{1+(\omega+\omega_0)^2\tau_0^2} + \dfrac{1-\omega_0(\omega-\omega_0)\tau_0^2}{1+(\omega-\omega_0)^2\tau_0^2}\right)$ | $\varepsilon_3''(\omega) = k/2$ $\left(\dfrac{\omega\tau_0}{1+(\omega+\omega_0)^2\tau_0^2} + \dfrac{\omega\tau_0}{1+(\omega-\omega_0)^2\tau_0^2}\right)$ |
| $\varepsilon_4'(\omega) - \varepsilon_\infty = k + k\omega a^2$ $\{\dfrac{e^{-a\omega}}{4a}\left(e^{\omega_0 a}Ei[a(\omega-\omega_0)] + e^{-\omega_0 a}Ei[a(\omega+\omega_0)]\right) -$ $\dfrac{e^{a\omega}}{4a}\left(e^{\omega_0 a}Ei[-a(\omega+\omega_0)] + e^{-\omega_0 a}Ei[a(\omega_0-\omega)]\right)\}$ | $\varepsilon_4''(\omega) =$ $k\dfrac{\pi a}{2}\omega\begin{bmatrix}e^{-\omega_0 a}\cosh a\omega & (\omega\leq\omega_0)\\ e^{-\omega a}\cosh a\omega_0 & (\omega\geq\omega_0)\end{bmatrix}$ |

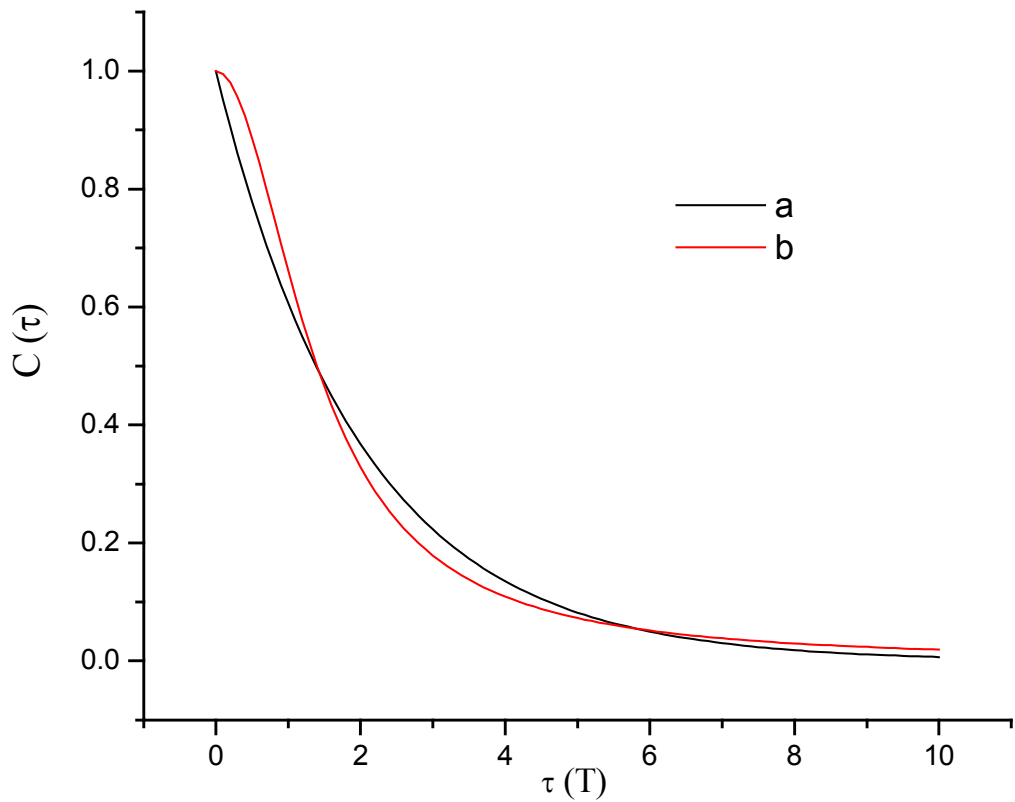

Fig.1. The exponential function (a) ($C_1(\tau)$, ($\tau_0 = 2$ T)) and the new function (b) ($C_2(\tau)$, (a=1.4 T). T represents an unspecified unit of dimension Time.

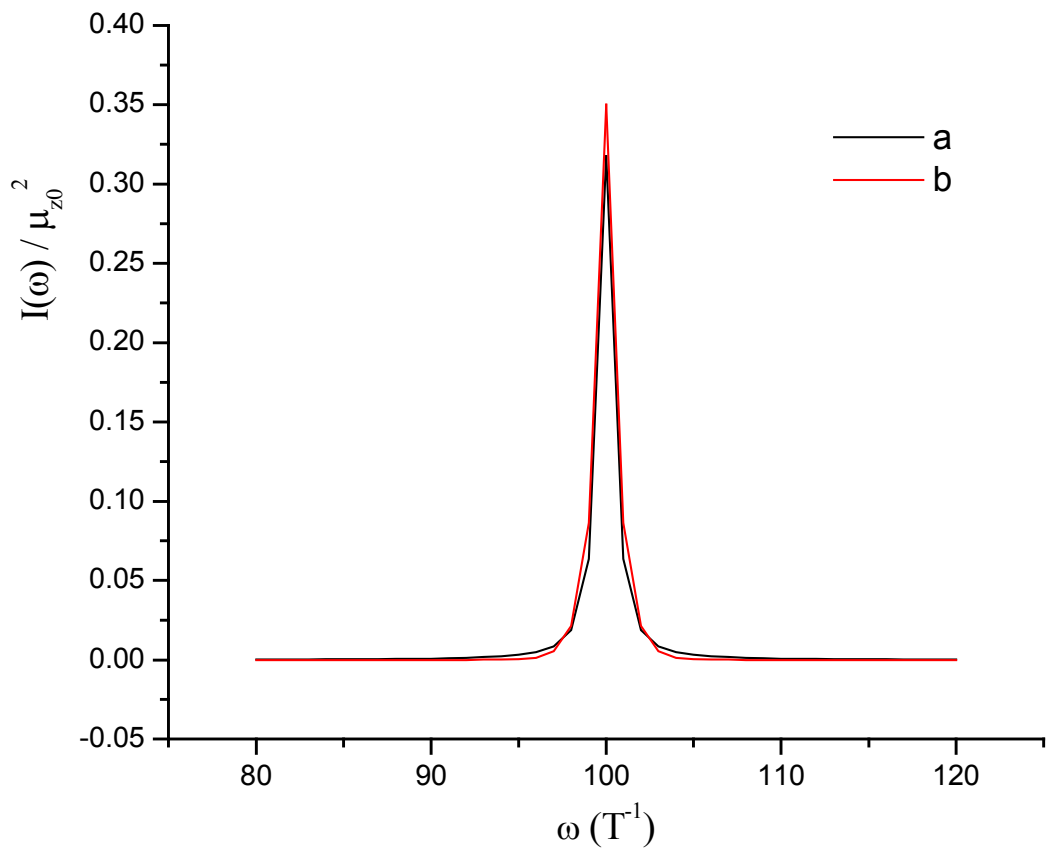

Fig.2. The Lorentzian lineshape, (a) with $\tau_0 = 2$ T and (b) the new lineshape function with a = 1.4 T. ($\omega_0 = 100$ T$^{-1}$).

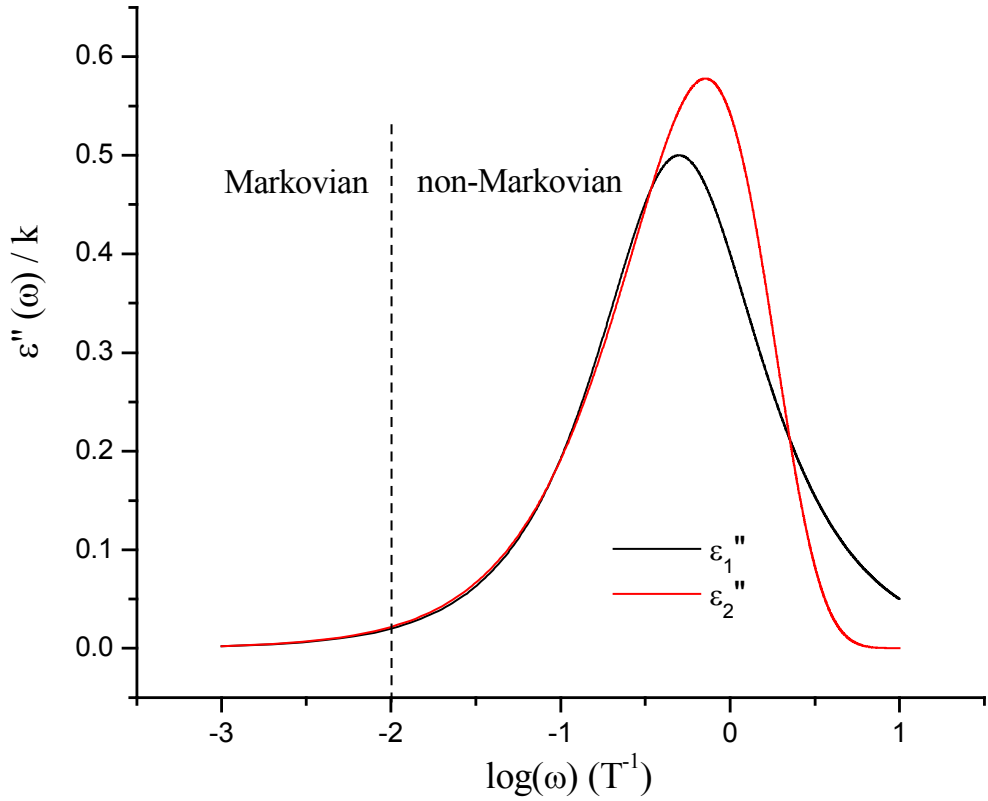

Fig.3. Imaginary part of the dielectric constants, (a) $\varepsilon_1''(\omega)$ from Debye theory with $\tau_0 = 2$ T and (b) $\varepsilon_2''(\omega)$ using the new time-correlation function with a = 1.4 T. Above frequency $\log(\omega) \approx -2.0$ T$^{-1}$ the process in non-Markovian and $\varepsilon_1''(\omega)$ is theoretically incorrect.

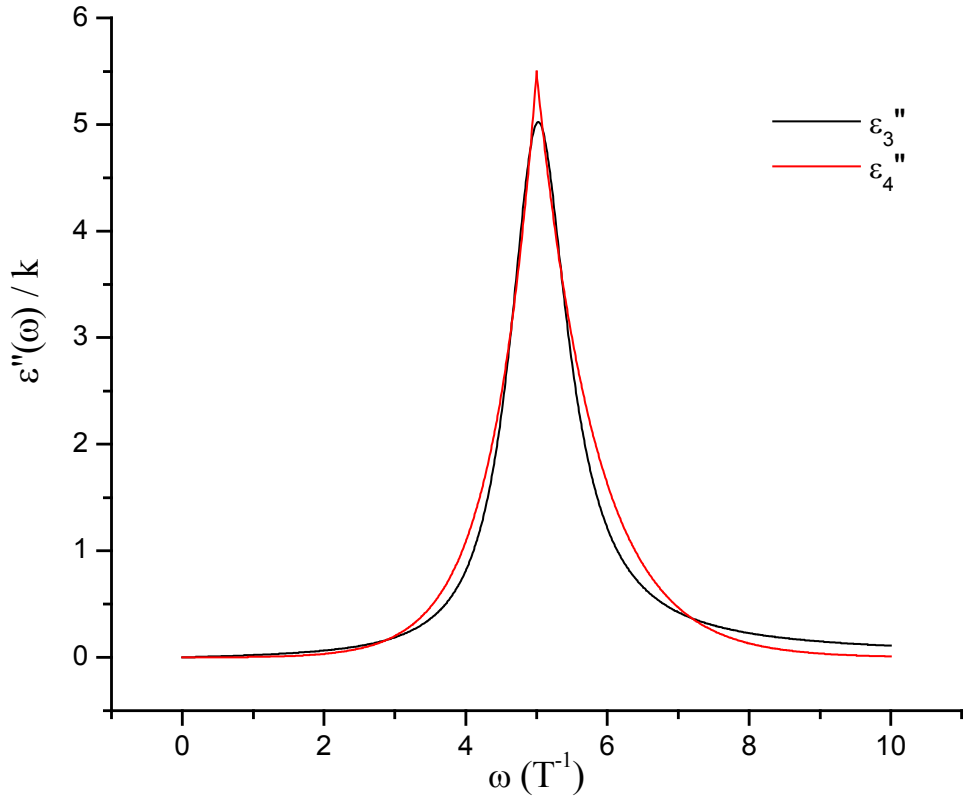

Fig. 4. Imaginary part of the dielectric constant (a) $\varepsilon_3''(\omega)$ obtained from Van Vleck-Weisskopf models with $\tau_0 = 2$ T and (b) $\varepsilon_4''(\omega)$ obtained using the new time-correlation function with a = 1.4 T. ($\omega_0 = 5$ T$^{-1}$)